\documentclass[aps,prb,preprintnumbers,amsmath,amssymb,superscriptaddress,twocolumn,10pt]{revtex4-2}

\pdfoutput=1

\usepackage{txfonts}
\usepackage{graphicx}
\usepackage{dcolumn}
\usepackage{bm}
\usepackage{array}
\usepackage[dvipsnames]{xcolor}
\usepackage[%
    colorlinks=true,
    pdfborder={0 0 0},
    linkcolor=blue,
    citecolor=black
]{hyperref}  
\usepackage{chngpage}
\usepackage{appendix}
\usepackage{subfig}
\usepackage{comment}
\usepackage{booktabs}
\usepackage{caption}
\newcommand{\ket}[1]{\lvert#1\rangle}

\newcommand{\be}{\begin{equation}}
\newcommand{\ee}{\end{equation}}
\newcommand{\bea}{\begin{eqnarray}}
\newcommand{\eea}{\end{eqnarray}}

\begin{document}

\title{Emergence of quadrupolar order under magnetic field
in  5$d^2$ double perovskites}
\author{Stashu Kozlowski}
	\affiliation{Department of Physics, University of Toronto, Toronto, Ontario M5S 1A7, Canada}
\author{Derek Churchill}
\affiliation{Department of Physics, University of Toronto, Toronto, Ontario M5S 1A7, Canada}
\author{Hae-Young Kee}
\email{hy.kee@utoronto.ca}
	\affiliation{Department of Physics, University of Toronto, Toronto, Ontario M5S 1A7, Canada}
	\affiliation{Canadian Institute for Advanced Research, CIFAR Program in Quantum Materials, Toronto, Ontario M5G 1M1, Canada}
\date{\today}

\begin{abstract}
Motivated by the time-reversal symmetry breaking signal in muon spin relaxation  below a transition temperature without accompanying noticeable magnetic Bragg peaks in $5d^2$ Os double perovskites, a rare ferro-octupolar order was proposed to account for such hidden order. 
Here we study the phase transitions under a magnetic field in triangular and face-centered cubic lattices using classical Monte Carlo simulations. It is expected that higher-rank moments do not couple linearly to the magnetic field. Consequently, a field applied along the ferro-octupolar order is not anticipated to influence the ordering. However, we observe the emergence of antiferro-quadrupolar ordering mixed with the antiferro-octupolar order (AFQO) due to the field-induced bond-dependent exchange interaction. This field-linear interaction among non-Kramer doublets arises via the coupling to the excited triplet states enabled by the external field. In the triangular lattice, we uncover intriguing vortex states, which could inspire future research into $5d^2$ triangular lattice systems. 
\end{abstract}

\maketitle

\section{Introduction}
In transition-metal Mott insulators, orbital degeneracy is often lifted by Jahn-Teller effects, where the crystal lattice distorts, separating the degenerate orbitals. This splitting is commonly observed in $e_g$ orbitals, where an octahedral crystal distorts into a tetragonal structure  \cite{Paolasini2002PRL, Margadonna2006ACS, Rodriguez1998PRB, Zolfaghari2013IOP, Naheed2024PRB} leading to the low-energy theory given by spin-only Hamiltonians \cite{Kugel1982}. However, when spin-orbit coupling (SOC) is strong and Jahn-Teller effects are weak, the spin and orbital degrees of freedom become intertwined and are described by the total angular momentum, known as pseudospin, $J$ \cite{Khal2005PTPS}. For a $d^2$ ion with strong SOC in an octahedral environment, it has been shown that the 
$J=2$ ground state can be further split due to $t_{2g}$ and $e_g$ mixing, resulting in a new ground state that can host both quadrupolar and octupolar moments  \cite{Maharaj2020PRL, Paramekanti2020PRB,Voleti2020PRB}.

While multipolar ordered phases have been widely explored and observed in $f$-electron compounds, such as NpO$_2$, PrX$_2$Al$_{20}$  (X = Ti, V), and others \cite{Santini2000PRL, Paixao2002PRL,Kiss2003, Kiss2005PRB, Tokunaga2006PRL, Sakai2011JPS, Suzuki1997JPS, Freyer2018PRB,Santini2009RMP, Onimaru2011PRL}, realizing a material with a $d$-orbital ion in an octahedral environment, requiring strong SOC and weak Jahn-Teller effects, has been challenging. This is because $d$-orbital systems typically exhibit weaker SOC and stronger Jahn-Teller distortions compared to their $f$-orbital counterparts \cite{Fazekas1999, Kugel1982}. However, a recent class of osmium (Os) double perovskites, Ba$_2$BOsO$_6$
(B = Ca, Mg) with a single transition \cite{Thompson2014JPCM, Marjerrison2016PRB, Maharaj2020PRL} has been claimed to exhibit octupolar moments \cite{Maharaj2020PRL, Voleti2020PRB, Paramekanti2020PRB, Pourovskii2021PRL}. In this structure, the Os$^{5+}$ (5$d^2$) ions form a face-centered cubic (FCC) lattice, with each ion confined in an octahedral crystal field.

To understand a microscopic route to possible octupolar orders, a microscopic Hamiltonian was developed that takes into account dominant intra- and inter-orbital exchange processes  \cite{Khaliullin2021PRR}. Through this model, bond-dependent interactions among quadrupolar moments via intra-orbital exchange were identified. The octupolar interaction was found to be antiferromagnetic, leading to frustration of the octupolar order on the FCC lattice, allowing quadrupolar order to emerge instead \cite{Khaliullin2021PRR}. However, when all intra- and inter-orbital hopping terms are included, it was suggested that this family of Os double perovskites lies at the boundary between a ferro-octupolar phase and a type-I antiferro-quadrupolar phase \cite{Churchill2022PRB,Voleti2021PRB}.

In a subsequent study expanding on the Hamiltonian including an external magnetic field, it was shown that an applied field induces an intriguing bond-dependent interaction between quadrupolar and octupolar moments, introducing a new form of frustration \cite{Rayyan2023PRB}. On the two-dimensional honeycomb lattice, this frustration enables the realization of a Kitaev spin liquid in multipolar systems \cite{Rayyan2023PRB}.

Here we investigate the field-induced phases of $d^2$ atoms with strong SOC in an FCC lattice, where each atom experiences octahedral crystal field splitting. Using classical Monte Carlo simulated annealing, we determine the classical ground states for various field directions. When the field is applied along the [111] body-centered direction, a ferro-octupolar (FO) moment is expected to be enhanced, as the magnetic field directly couples to the octupolar moment, similar to a Zeeman-like term. However, this natural expectation breaks down when the bond-dependent interaction induced by the field becomes dominant over the Zeeman-like coupling, leading to the emergence of an antiferro-quadrupolar ordering mixed with the antiferro-octupolar order (AFQO). Since the FCC lattice can be viewed as a stack of triangular lattices, we also examine the phase diagram in the triangular lattice, and find the emergence of intriguing vortex states that mix quadrupolar and octupolar configurations.

The paper is structured as follows: in Section 2, we provide an overview of the double perovskite structure and the FCC lattice of osmium atoms, along with a review of recent experimental and theoretical developments. In Section 3, we derive the field-induced Hamiltonian for the triangular and FCC lattices. Our phase diagrams for the triangular and FCC lattices are presented in Section 4. The discussion and summary are provided in the final section.

\section{Review on double perovskites and pseudospin model}

\subsection{ Double perovskites and the underlying FCC lattice}

Double perovskite compounds, A$_2$ BB'O$_6$ are derived from simple perovskites, ABO$_6$, but with half of the B atoms replaced with B' atoms such that any two nearest neighbour (n.n.) B-sites exclusively contain only one B and B' atom. The end effect being the formation of two FCC lattices that intersect with each other, with one lattice being entirely composed of B and the other of B' atoms as shown in Fig. 1(a). The B and B' positions are filled by transition metal elements, with the oxygen atoms forming octahedral cages around them. For Ba$_2$BOsO$_6$ with B = Ca, Cd, Mg, B atoms
are non-magnetic leading to a FCC lattice of $d^2$ multiplets \cite{Churchill2022PRB}.

Two different coordinate systems are commonly used for describing spin exchange interactions in the literature, so we begin by reviewing these two bases and their relationship. Figures 1(b) and 1(c) depict the octahedral and crystallographic orthonormal bases used for constructing the FCC B' lattice. The relationship between the two bases is given by the following:
\begin{equation}
    \hat{\mathbf{a}} = \frac{1}{\sqrt{6}}\left(\hat{\mathbf{x}} + \hat{\mathbf{y}} -2\hat{\mathbf{z}} \right), \; \hat{\mathbf{b}} = \frac{1}{\sqrt{2}}\left(\hat{\mathbf{y}} - \hat{\mathbf{x}}\right),\;\hat{\mathbf{c}} = \frac{1}{\sqrt{3}}\left( \hat{\mathbf{x}} + \hat{\mathbf{y}} + \hat{\mathbf{z}}\right),
\end{equation}
where the vectors ${\hat x}$, ${\hat y}$ and ${\hat z}$ are defined based on the positions of oxygen atoms, while ${\hat a}$ and ${\hat b}$ and ${\hat c}$ are aligned with the crystallographic coordinates.

Throughout the paper, we will use the notation $\left[\alpha\beta\gamma \right]$ to denote a direction in the basis of the octahedral lattice vectors. For example, $\left[ 111\right]$ denotes the same direction as $\hat{\mathbf{x}} + \hat{\mathbf{y}} + \hat{\mathbf{z}}$. A line over a number indicates the opposite direction. For example, $\left[ \Bar{1} 1 0\right]$ is the same direction as $\hat{\mathbf{y}}-\hat{\mathbf{x}}$. 
Note that the FCC lattice has a
$\mathcal{C}_3$ rotational symmetry about the $[111]$ direction and can be considered as a stacked triangular lattice along the $c$-axis.

\begin{figure}[]
    \centering
    \includegraphics[width=\linewidth]{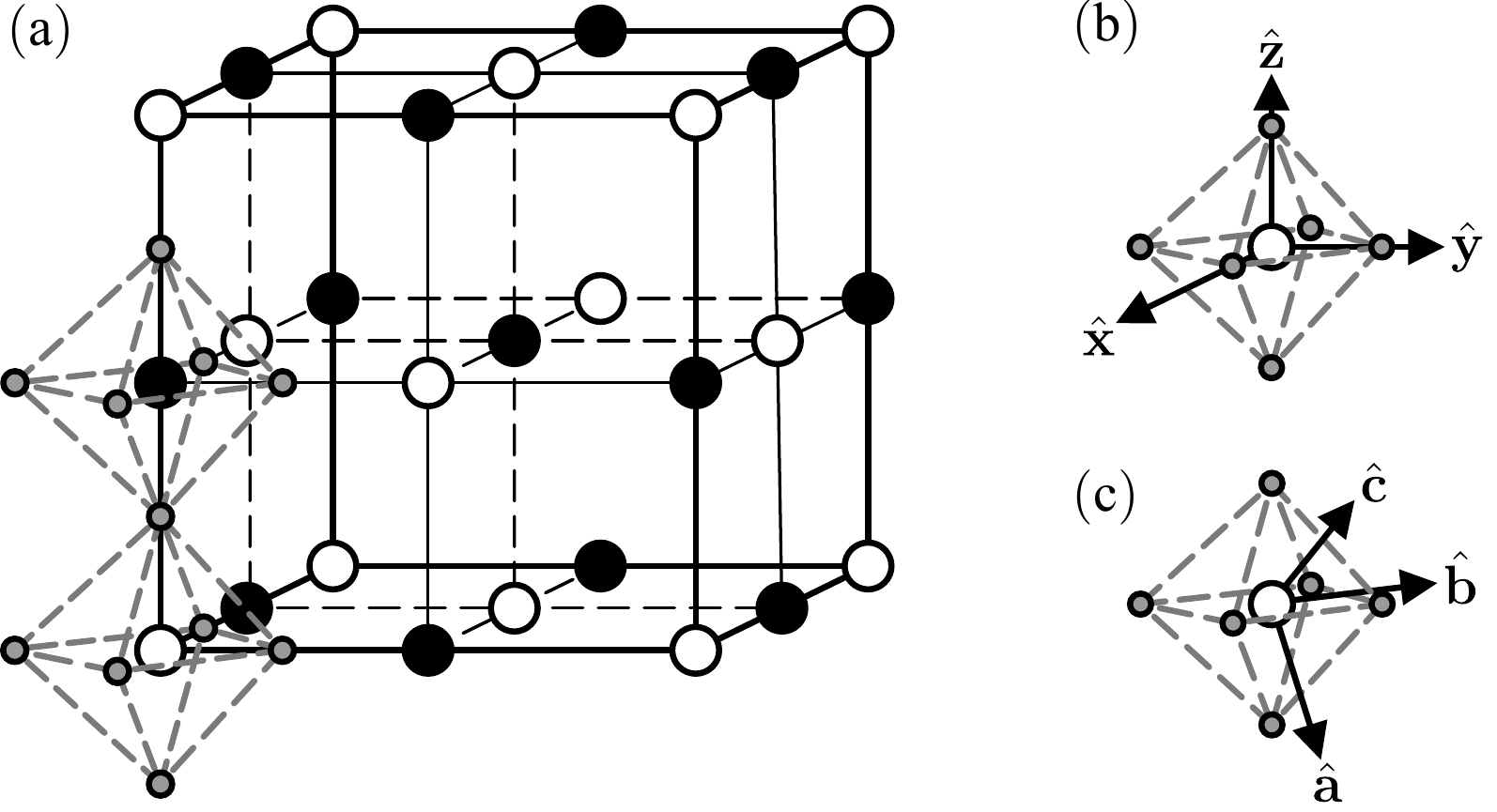}
    \caption{(a) Double perovskites with an interlocking FCC lattice formed by B and B' ions, represented as white and black spheres, respectively, with electrons depicted in gray. For clarity, only two octahedral oxygen structures are shown. (b) The octahedral coordinate system $\hat{\mathbf{x}}$, $\hat{\mathbf{y}}$, and $\hat{\mathbf{z}}$. (c) The crystallographic axis $\hat{\mathbf{a}}$, $\hat{\mathbf{b}}$, and $\hat{\mathbf{c}}$.}
    \label{fig:mesh1}
\end{figure}

\subsection{Osmium double perovskites}

Various experimental and theoretical investigations have been conducted in Os-based double perovskites, Ba$_2$BOsO$_6$ (B = Ca, Mg), by various groups. Experimental investigations of this family of perovskites have found a peak in their heat capacity at $T^* \sim 50 $ K, suggesting a phase transition \cite{Thompson2014JPCM, Marjerrison2016PRB}. Measurement of magnetic susceptibility and fitting of inverse susceptibility to the Curie-Weiss law provide a negative Curie-Weiss temperature of $\Theta_{CW} \sim - 130$ K, commonly associated with type-I antiferromagnetic order \cite{Thompson2014JPCM, Marjerrison2016PRB}. Similarly, inelastic neutron diffraction experiments observed strong, gapped, magnetic spectral weight at wavevectors [100] $(|Q|^{-1} \sim 0.78 \; \AA^{-1})$ and [110] $(|Q|^{-1} \sim 1.1 \; \AA^{-1})$, typical of the antiferromagnetic order  \cite{Maharaj2020PRL}. However, further neutron powder diffraction investigations revealed an absence of Bragg scattering at low temperatures at the [100] and [110] positions, contrary to the expectation  \cite{Maharaj2020PRL} making it a hidden order.  Neutron powder diffraction experiments reveal a dipole ordered moment of $0.06\mu_B$ to $0.13 \mu_B$, depending on the type of material substituted into the $B$ atom  \cite{Maharaj2020PRL}. Lastly, low temperature $\mu$SR experiments show an oscillating asymmetry over time, indicating the presence of internal magnetic fields and thus time reversal symmetry breaking  \cite{Thompson2014JPCM, Marjerrison2016PRB}.

Furthermore, subsequent x-ray diffraction experiments have shown no changes in crystal symmetry  \cite{Maharaj2020PRL}, indicating the absence of quadrupolar order, as it would lower the crystal symmetry from cubic to tetragonal  \cite{ Paramekanti2020PRB, Maharaj2020PRL, Voleti2020PRB}.
Thus, another proposal to explain the current data was developed.
As briefly mentioned in the Introduction, treating the $t_{2g}-e_g$ mixing as a perturbation, the five-fold degenerate $J=2$ state is broken into a lower non-Kramers doublet ground state and an excited triplet  \cite{Voleti2020PRB, Maharaj2020PRL, Paramekanti2020PRB}. In resemblance with the crystal field states $e_g$ and $t_{2g}$, the lower non-Kramers doublet and excited triplet were referred to as $E_g$ and $T_{2g}$, respectively  \cite{Rayyan2023PRB, Churchill2022PRB, Takayama2021JPS, Khaliullin2021PRR}. The lower $E_g$ state does not carry dipole moment, but rather higher moments such as quadrupolar and octupolar \cite{Taka2000JPSJ,Maezono2000PRB,Brink2001PRB}.
It was proposed that the ground state exhibits an FO order \cite{Voleti2020PRB}, which leads to
an internal orbital current and hence a magnetic field within the crystal, consistent with $\mu$SR data.
However, since the muon might impact the local crystal potential leading to a revival of dipole moment \cite{Gomilsek2023CP, Voleti2023npj}, further investigation of the hidden order needs to be explored. Furthermore, it has been challenging to experimentally detect a quadruoplar order  \cite{Mydosh2011RMP, Patri2019Nature, Sorn2024PRB}, and thus a quadrupolar order has not been completely ruled out.

\subsection{Local Hamiltonian and non-Kramer's doublet}
Motivated by the above results, a microscopic model for the $E_g$ non-kramer doublet was derived to understand the exchange processes leading to the octupolar order and its competition with the quadrupolar orders  \cite{Khaliullin2021PRR,Churchill2022PRB}.
Here we briefly review how to derive a microscopic Hamiltonian for the multipolar moments.

To construct an exchange model, one begins with a local atomic site $i$ Hamiltonian, $H_i$. After projecting out the high energy $e_g$ states due to the large octahedra crystal field splitting, $H_i$ is given by \cite{Churchill2022PRB, Rayyan2023PRB},
\begin{multline}
    H_i = U \sum_m n_{m +} n_{m-} + U' \sum_{m \neq m'} n_{m+} n_{m'-} + (U' - J_H) \sum_{m < m', \sigma } n_{m\sigma}n_{m' \sigma}  \\ + J_H \sum_{m \neq m'} (c^\dagger_{m+}c^\dagger_{m'-} c_{m-} c_{m'+} + c^\dagger_{m+} c^\dagger_{m-} c_{m'-} c_{m'+}) + \lambda \;\mathbf{L}_\text{eff}\cdot \mathbf{S},
\end{multline}
where the summation is performed over orbitals exclusively in the $t_{2g}$ basis and $\mathbf{L}_\text{eff} = -1$ due to the projecting out the $e_g$ orbitals.

Using Hund's rules and SOC, one finds a five-fold degenerate $J=2$ as the lowest state. However, $e_g-t_{2g}$ mixing causes $J=2$ to split, and such splitting of the $J=2$ multiplet can be modeled by the cubic crystal field Hamiltonian given by  \cite{Voleti2020PRB, Maharaj2020PRL, Paramekanti2020PRB}, 
\begin{equation}
    H' = \delta (\mathcal{O}_4^0 + 5\mathcal{O}_4^4),
\end{equation}
where $\mathcal{O}_4^0$ and $\mathcal{O}_4^4$ are Stevens operators given by \cite{Paramekanti2020PRB, Maharaj2020PRL}
\begin{eqnarray}
    \mathcal{O}_4^0 &= & 35J_z^4 - \left[ 30J\left(J + 1\right)-25\right]J_z^2 + 3J^2\left(J+1\right)^2 -6J\left(J+1\right),\nonumber \\
    \mathcal{O}_4^4 &=& \frac{J_+^4 + J_-^4}{2}.
\end{eqnarray}
Diagonalizing $H'$ yields the $E_g$ ground state, expressed in $\ket{J^z}$ states as
\begin{equation}
    \ket{\uparrow} = \frac{1}{\sqrt{2}} \left( \ket{2} + \ket{-2}\right), \;\;\; \ket{\downarrow} = \frac{1}{\sqrt{6}} \ket{0}.
\end{equation}

Using the $E_g$ wavefunctions, the matrix representations of the two quadrupole and one octupole operators respectively are given as  \cite{Churchill2022PRB}
\begin{equation}
    Q_{x^2-y^2} = J_x^2 - J_y^2, \;\;\;Q_{3z^2-r^2} = \frac{3J_z^2 - J^2}{\sqrt{3}}, \;\;\; T_{xyz} = \frac{\sqrt{15}}{6} \overline{J_xJ_yJ_z},
\end{equation}
where the overline symbol denotes symmetrization of the underlying operators. They can be expressed as Pauli  matrices as follows \cite{Rayyan2023PRB, Churchill2022PRB}:
\begin{equation}
    s^1 \equiv \frac{1}{4\sqrt{3}} Q_{x^2-y^2}, \;\;\; s^2 \equiv \frac{1}{6\sqrt{5}} T_{xyz},\;\;\; s^3 \equiv \frac{1}{4\sqrt{3}} Q_{3z^2-r^2}.
\end{equation}
It is important to highlight that the quadrupole and a octupole operators 
are defined in the crystallographic basis shown in Fig. 1(c). As such, we will use the notation $(s^a, s^b, s^c) = (s^3, s^1, s^2)$ as effective pseudospin-1/2 operators. Under such a system, the quadrupolar components, $s^a$ and $s^b$, are in a plane perpendicular to $[111]$, while the octupolar component, $s^c$ is parallel to the $[111]$ line, i.e, $c$-axis.

\subsection{Nearest neighbour exchange Hamiltonian}
To construct the exchange Hamiltonian, we begin by considering the hopping processes between two sites, $i$ and $j$, and then apply the lattice symmetry to generalize the exchange interactions for the remaining bonds. Let us consider a tight-binding model on the line spanned by $\hat{\mathbf{b}}$ (see Fig. 2(a)), which is referred to as a $z$-bond since the dominant hopping is of the form $xy \leftrightarrow xy$ and hence are perpendicular to the $z$-axis.

Denoting the n.n. sites as $\langle ij\rangle_z$ (the subscript indicates the bond type), the three types of orbital hopping integrals is given by
\begin{figure}[]
    \centering
    \includegraphics[width = \linewidth]{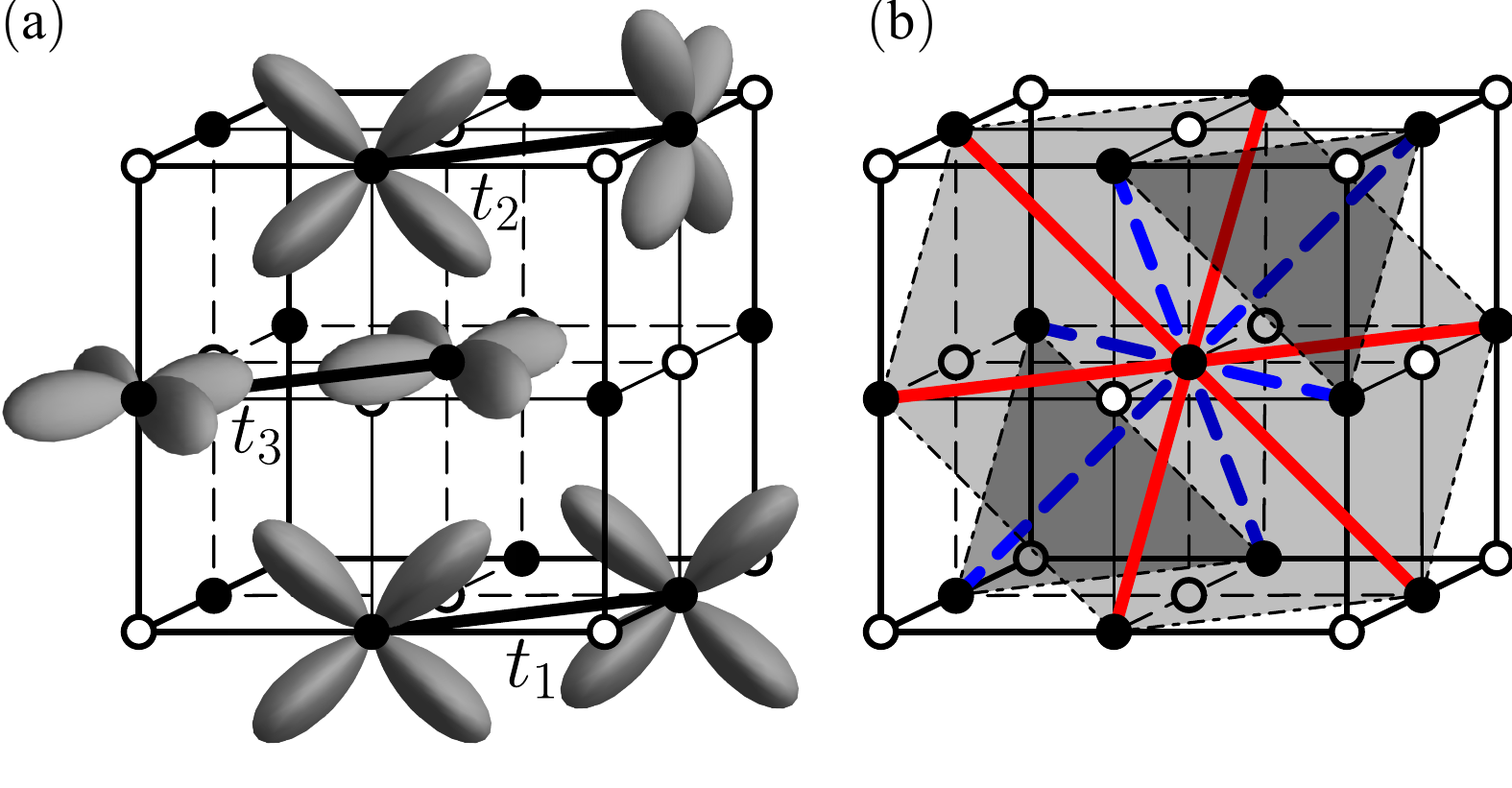}
    \caption{(a) Examples of exchange processes among $t_{2g}$ orbitals on the FCC lattice: $t_1$ corresponds to hopping between $zy$ and $zy$ orbitals, $t_2$ to hopping between $yz$ and $xz$ orbitals, and $t_3$ to hopping between two $xy$ orbitals. (b) Dominant hopping within and between the planes perpendicular to $[111]$ is illustrated. Solid red lines represent in-plane bonds (1–6), while dashed blue lines indicate out-of-plane bonds (7–12). Specifically, bonds 2, 5, 9, and 12 are $z$-bonds; bonds 3, 6, 8, and 11 are $x$-bonds, and bonds 1, 4, 7, and 10 are $y$-bonds.}
    \label{fig:mesh2}
\end{figure}
\begin{equation}
    H_{\langle ij \rangle_z}^\text{TB} = c_i^\dagger T_{ij} c_j + c_j^\dagger T_{ij}^\dagger c_i,
\end{equation}
where $c_i^\dagger = \left(c_{i, xy}^\dagger, c_{i, xz}^\dagger, c_{i, yz}^\dagger\right)$  \cite{Churchill2022PRB} 
and
\begin{equation}
        T_{ij}^z = \begin{pmatrix}
        t_3 & 0 & 0 \\
        0 & t_1 & t_2 \\
        0 & t_2 & t_1
    \end{pmatrix}.
\end{equation}

Using the strong coupling expansion, the exchange Hamiltonian between the non-Kramer doublets along the $z$-bond takes on the following form \cite{Churchill2022PRB}:
\begin{equation}
    H_{\langle ij\rangle_z} = J_\tau s^a_i s^a_j +J_Q \left(s^a_i s^a_j + s^b_i s^b_j \right)+J_O s^c_i s^c_j 
\end{equation}
where to second order the exchange parameters are given as \cite{Churchill2022PRB}
\begin{eqnarray}
    J_\tau &= & \frac{4}{9} \frac{(t_1 - t_3)^2}{U}, \; J_Q = \frac{2}{3} \frac{t_1(t_1+2t_3)-t^2_2}{U}, \nonumber\\
    J_O & =& \frac{2}{3} \frac{t_1 (t_1 + 2t_3) +t^2_2}{U},
\end{eqnarray}
where Hund's coupling has been ignored.
Note that when $t_1 =0$, $J_O (= - J_Q) > 0$, leading to frustration of the octupolar moment on the FCC lattice, consistent with the Hamiltonian reported in Ref.  \cite{Khaliullin2021PRR}. However, when $t_1$ becomes finite, the octupolar interaction $J_O$ can become negative, potentially leading to an octupolar order \cite{Churchill2022PRB}.

The exchange interactions for the remaining bonds can be determined by applying the lattice symmetry. Below, we present the complete Hamiltonian for the triangular and FCC lattices, incorporating the effects of an external magnetic field.

\section{Hamiltonian including external magnetic field for multipolar moments in triangle and FCC lattices}

\subsection{Triangular lattice}

Now, let us consider the Zeeman term given by $H_i^z = \mu_B (\mathbf{L}_i + 2\mathbf{S}_i)\cdot \mathbf{h}$, where  $\mathbf{h} = h_x \hat{\mathbf{x}} + h_y \hat{\mathbf{y}} + h_z \hat{\mathbf{z}}$.
Based on symmetry considerations, multipolar moments do not couple to the field linearly. Instead we expect terms like $(2h_z^2 - h_x^2 - h_y^2) s^z$, $(h_x^2 - h_y^2) s^b$ and $(h_x h_y h_z) s^c$.
However, what is overlooked in the symmetry analysis are the virtual processes involving the higher-energy triplet $T_{2g}$, separated by $\Delta$.
Since $T_{2g}$ carries dipole moments, it is influenced by the Zeeman term. This splitting can induce additional exchange interactions between the quadrupolar and octupolar moments, and a linear Zeeman-like term may also emerge. We will now explore the full Hamiltonian, incorporating the external field in the triangular lattice.

Following the strong coupling approach used for the honeycomb lattice  \cite{Rayyan2023PRB}, we derive the exchange terms for the triangular lattice, assuming $h/\Delta$ is a small parameter, alongside $t_i/U$. Let's first consider the $z$-bond. The additional quadratic terms take the following form:
\begin{equation}
    H^B_{\langle ij\rangle_z} =  - \sqrt{2}J_B^z (s^a_i s^c_j + s^c_i s^a_j),
\end{equation}
where 
\begin{equation}
    J_B^z  =  \frac{8}{3\sqrt{6}} \frac{t_2(2t_1+t_3)}{U} \frac{g_J \mu_B h^z}{\Delta}.
\end{equation}
Since this term involves virtual hopping processes to the triplet 
$T_{2g}$, the coefficient $J_B^z$
is proportional to $h_z/\Delta$, where $\Delta$ represents the energy splitting between $T_{2g}$ and $E_g$.
Note that $J_B^z$ is only finite when the magnetic field has a $z$-component.

The remaining exchange interactions for $\langle ij\rangle_y$ (n.n. $y$-bonds) and $\langle ij\rangle_x$ (n.n. $x$-bonds) within the triangular lattice perpendicular to the $[111]$-axis as shown in Fig. 2(b) can be obtained by performing a $\mathcal{C}_3$ rotation about the $[111]$-axis. Under this rotation, we note that the spin operators transform as follows:
\begin{equation}
    s^a \to -\frac{1}{2}s^a + \frac{\sqrt{3}}{2}s^b, \;\;\; s^b \to -\frac{\sqrt{3}}{2}s^a - \frac{1}{2}s^b, \;\;\; s^c \to s^c.
\end{equation}
Rotating once will give us the quadratic $x$-bond and once more after will give the $y$-bond. 

Lastly, $\mathbf{h}_\text{eff} = h_\text{eff}^a \hat{\mathbf{a}} + h_\text{eff}^b \hat{\mathbf{b}} + h_\text{eff}^c \hat{\mathbf{c}}$ is expressed in crystallographic basis and is given by
\begin{equation}
\begin{split}
h_\text{eff}^a &= 6(g_J\mu_B)^2 \frac{2h_z^2 - h_x^2 - h_y^2}{\Delta} \\
h_\text{eff}^b &= 6(g_J\mu_B)^2 \frac{h_x^2 - h_y^2}{\Delta} \\
h_\text{eff}^c &=\frac{4}{3\sqrt{3}} \frac{t_2(t_1 - t_3)g_J \mu_B}{U} \frac{(h_x + h_y + h_z)}{\Delta}
- 36\sqrt{3} (g_J\mu_B)^3 \frac{h_xh_yh_z}{\Delta^2}.
\end{split}
\end{equation}
The above forms of $h_{\rm eff}$ are expected based on the symmetry.
The resulting Hamiltonian is same as the one obtained for the honeycomb lattice \cite{Rayyan2023PRB}, except that there are six n.n. bonds with two x-, two y-, and two z-bonds in the triangular lattice.

\subsection{FCC lattice}

For the FCC lattice, there are 12 bonds connecting to a site.
6 bonds denoted by the number 1 to 6 (denoted by red line) as shown in Fig. 2(b)  are within the triangular plane discussed in the above subsection, while the remaining 6 bonds denoted by the number 7 to 12 (denoted by blue line), connect sites between the n.n. belong to different triangles, i.e., out-of-plane bonds.
For the bonds within the plane (red bonds), the tight binding parameter is the same as Eq. (9).
However, the 6 bonds connecting outside the triangular plane (blue bonds) have different tight binding parameters. For example the $z$-bond along $[110]$-direction takes the form of
\begin{equation}
        T_{\langle ij \rangle \in {\rm out-of-plane}}^z= \begin{pmatrix}
        t_3 & 0 & 0 \\
        0 & t_1 & -t_2 \\
        0 & -t_2 & t_1
    \end{pmatrix},
\end{equation}
where the sign of $t_2$ changes due to the bond direction.
This leads to $J_B$ interactions gaining a negative sign, while the other terms are not altered as summarized below.

Combining all the exchange contributions, the full effective Hamiltonian for the FCC lattice can then be expressed as:
\begin{equation}
    H = \sum_{ \langle ij \rangle_\gamma} J_\tau \tau_i^\gamma \tau_j^\gamma  +J_Q \left(s^a_i s^a_j + s^b_i s^b_j \right)+J_O s^c_i s^c_j+H^{B}_{{\langle ij \rangle}_\gamma}- \sum_i \mathbf{h}_\text{eff} \cdot \mathbf{s}_i,
\end{equation}
where
\begin{eqnarray}
    H^{B}_{\langle ij \rangle_\gamma} =
    \begin{cases} 
     {\color{blue} -} \sqrt{2}J_{B}^\gamma (\tau_i^\gamma s^c_j + s^c_i \tau_j^\gamma) & \langle ij \rangle_\gamma \in \text{in-plane} \\ 
    {\color{blue} +} \sqrt{2}J_{B}^\gamma (\tau_i^\gamma s^c_j + s^c_i \tau_j^\gamma)& \langle ij \rangle_\gamma \in \text{out-of-plane}
    \end{cases}
\end{eqnarray}
where $\gamma$ can take on values of $z,x,$ or $y$ and is determined by the type of bond $\langle ij \rangle$ (see Fig. 2(b)), $\tau_i^\gamma \equiv s^a \cos{\theta_\gamma} + s^b \sin{\theta_\gamma}$, where $\theta_\gamma = 0, \frac{2\pi}{3}, \frac{4\pi}{3}$ when $\gamma = z,x,y$. 
Note the sign difference in front of the $J_B$ term between the in-plane vs. out-of-plane bonds.

Furthermore, $\mathbf{h}_\text{eff} = h_\text{eff}^a \hat{\mathbf{a}} + h_\text{eff}^b \hat{\mathbf{b}} + h_\text{eff}^c \hat{\mathbf{c}}$ is expressed in crystallographic basis and is given by
\begin{equation}
\begin{split}
h_\text{eff}^a &= 6(g_J\mu_B)^2 \frac{2h_z^2 - h_x^2 - h_y^2}{\Delta} \\
h_\text{eff}^b &= 6(g_J\mu_B)^2 \frac{h_x^2 - h_y^2}{\Delta} \\
h_\text{eff}^c &= 
- 36\sqrt{3} (g_J\mu_B)^3 \frac{h_xh_yh_z}{\Delta^2}.
\end{split}
\end{equation}

The linear $h$ term observed in the honeycomb and triangular lattices with the form of 
$h_{\rm eff}^c= \frac{4}{3\sqrt{3}} \frac{t_2(t_1 - t_3)g_J \mu_B}{U} \frac{(h_x + h_y + h_z)}{\Delta}$ does not occur in the FCC lattice, because there are positive contributions from the n.n. in-plane (1-6)  and negative contributions from the n.n. out-of-plane (7-12). They cancel out.

The FCC lattice Hamiltonian does not follow the form as the one derived for the honeycomb and triangular lattices. In the FCC lattice, there are 12 bonds, and $J_B^\gamma$ alternates it sign, and the linear $h$ term in $h^c_{\rm eff}$ does not occur in the FCC lattice.

\section{Phase diagram and Field-induced phases}
Exchange parameters $J_\tau$, $J_Q$ and $J_O$ for Os double perovskites, Ba$_2$BOsO$_6$ (B = Ca, Cd, Mg) are reported using the combination of density functional theory and strong coupling expansion \cite{Churchill2022PRB}. As reported in Ref. \cite{Churchill2022PRB}, the exchange parameters, in the absence of an applied field, result in a FO ground state for Ba$_2$MgOsO$_6$, whereas Ba$_2$CaOsO$_6$ and Ba$_2$CdOsO$_6$ exhibit antiferro-quadrupole (AFQ) ground states. In all three cases, slight modifications to the exchange parameters can shift the ground state from FO to AFQ or vice versa. In light of this, our starting point is finding where this transition occurs. This is done classically by minimizing the Hamiltonian via Monte Carlo simulated annealing \cite{Kirkpatrick1983Science, MetropolisJCP1953, Kirkpatrick1984JSP, Hayami2024DAS}.

For fixed $J_\tau = 2.8$ and $J_Q = -1.5$, close to the exchange parameters for Os double perovskites (all exchange parameters are in unit of meV), we find that the phase transition from FO and AFQ occurs at approximately $J_O = -0.85$. 
To explore the effects of the magnetic field through $J_B$ and $h_{\rm eff}$ on FO and AFQ orders, we choose two different values of $J_O = -0.9$ and $J_O = -0.7$, as these values are near the phase transition and yield FO and AFQ phases, respectively.

Applying a field in the $[111]$-direction, where $\mathbf{h}_{\rm eff} = h^c_{\rm eff} \hat{\mathbf{c}}$, we observe that $J_B^x=J_B^y=J_B^z$. For clarity, let us define $J_B^\gamma \equiv J_B$ and $h_{\rm eff}^c \equiv h_{\rm eff}$. For Os double perovskites
the exchange parameters produce $J_B < 0$ \cite{Churchill2022PRB}. In order to explore the phase space, we will treat $J_B$ and $h_{\rm eff}$ as independent parameters, and present the phase diagram as a function of $J_B$ and $h_{\rm eff}$.
Before we present the classical phase diagram for the FCC lattice for Os double perovskites, we present the result on a triangular lattice first.

\subsection{Triangular lattice}
\begin{figure}[]
    \centering
    \includegraphics[width=\linewidth]{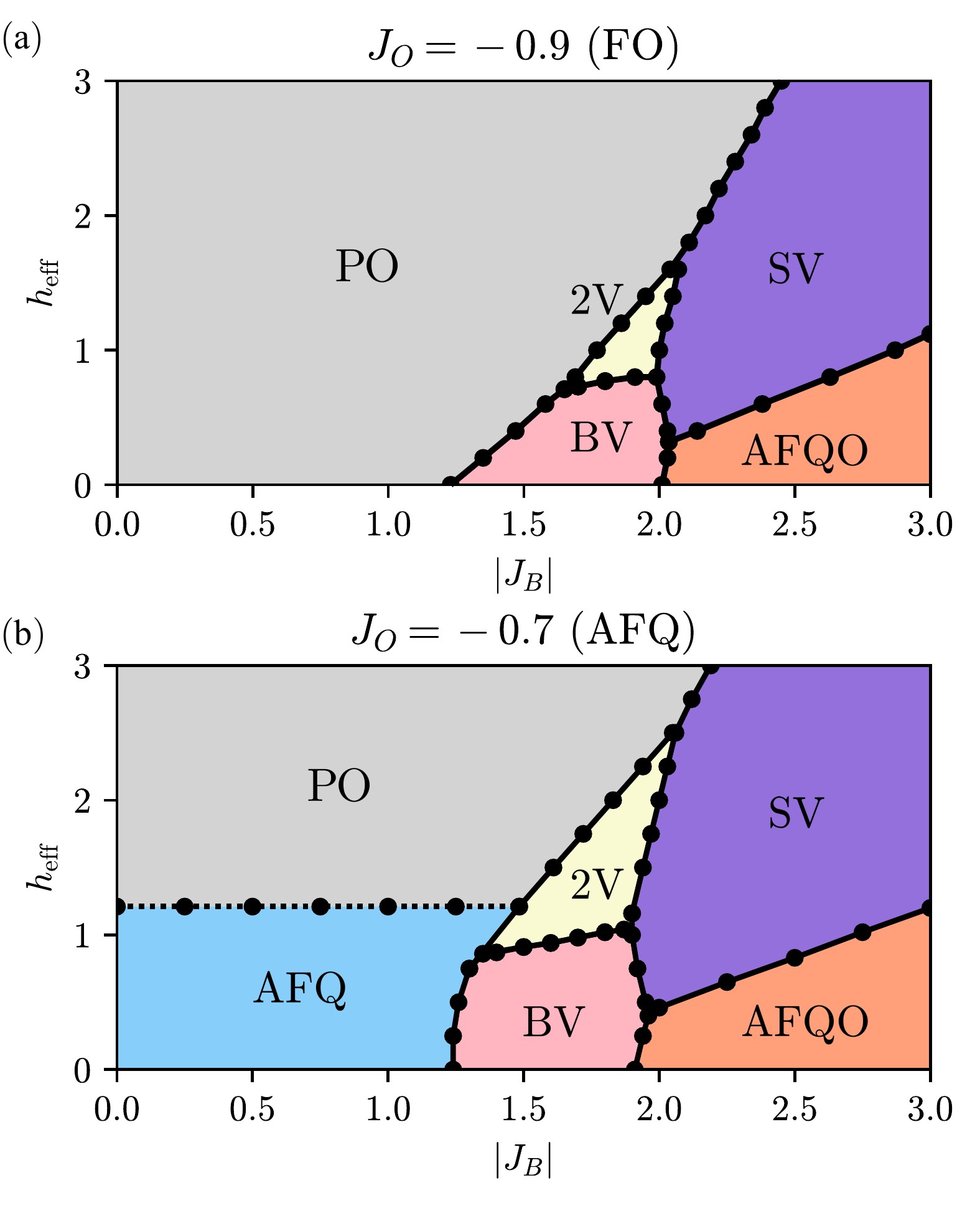}
    \caption{Classical phase diagrams computed by Monte Carlo simulation on the triangular lattice when a field is applied in the $[111]$-direction and (a) $J_O = -0.9$, and (b) $J_O = -0.7$. Solid and dashed lines represent first and second order phase transitions, respectively. Note that the ferro-octupolar (FO) phase at $h_{\rm eff}=0$ in (a) becomes the polarized octupolar (PO) state at $h_{\rm eff}\neq 0$ as the moment direction is fixed by the external field.}
    \label{fig:mesh3}
\end{figure}

Using $J_\tau =2.8 $, and $J_Q = -1.5$, 
we present the classical phase diagram obtained through Monte Carlo simulated annealing for the FO ($J_O = -0.9$) and AFQ ($J_O = -0.7$) cases in Fig. 3(a) and (b), respectively. Starting with the $J_O = -0.9$ case, we note that as the field $h_{\rm eff}$ and $J_B$ increases, there are several phases found. Starting from the FO phase, the effective $c$-axis field does not generate any transition, as expected. On the other hand, for $h_{\rm eff}=0$,  $J_B$ induces a field-induced vortex phase with a large unit cell (denoted by BV), which then transitions to AFQO phase as $J_B$ further increases. In the BV phase, the pattern of the quadrupole and octupole moment is shown in Fig. 4(a), where the unit cell of the pseudospin configurations show the staggered chiral vortex pattern. The AFQO phase is also shown in Fig. 4(c), where the quadrupole and octupole moment order are in a stripy antiferro pattern.

When both $h_{\rm eff}$ and $J_B$ are finite, the two other vortex phases are found. Starting in the BV phase with sufficiently large $J_B$ and increasing $h_{\rm eff}$, we see a transition into a state composed of two vortex (2V) alignments interacting in a anticyclonic manner as shown Fig. 4(b). Similarly, increasing $h_{\rm eff}$ while within the AFQO phase, we observe a transition into another vortex  with a smaller unit cell of a honeycomb (denoted as SV) as shown in Fig. 4(d). 

For $J_O = -0.7$, we observe similar phases, including AFQO, BV, PO, SV, and 2V. Starting with no applied field in the AFQ phase, there is no pseudospin octupolar component. However, as the magnitude of $h_{\rm eff}$ increases, a uniform rise in the octupolar component is observed at all sites, culminating in a second-order phase transition to the PO phase at $h_{\rm eff} = 1.2$. In a manner similar to the triangular $J_O = -0.9$ phase diagram, $J_B$ induces transitions to the BV and AFQO phases when no field is applied. Similarly, increasing $h_{\rm eff}$ from the BV or AFQO phases results in transitions to the 2V and SV phases, respectively.

As shown in Fig. 4, the BV and SB pseudospin configurations are invariant under $\mathcal{C}_3$ rotation. In contrast, the 2V and AFQO configurations are asymmetric with respect to this rotation. Consequently, the 2V and AFQO states are each degenerate with two additional pseudospin configurations generated by applying the $\mathcal{C}_3$ rotation to the depicted states.

The presence of the AFQO phase for low $h_{\rm eff}$ and large $J_B$ can by explained by examining the $J_B^\gamma$ terms in the full effective Hamiltonian. If the bond between two n.n. sites is a $z$-bond, the Hamiltonian term is proportional to $(s_i^a s_j^c + s_i^c s_j^a)$. This is minimized when $\mathbf{s}_i = \mathbf{s}_j = \pm(1,0,-1)/\sqrt{2}$. However, when considering a $x$- or $y$-bound, we see that the Hamiltonian is instead minimized when $\mathbf{s}_i = -\mathbf{s}_j = \pm (1,0,-1)/\sqrt{2}$. Put more succulently, if the n.n. bond is a $z$-bond, it favors a parallel pseudospin configuration with quadrupole and octupole components, while $x$- and $y$-bonds favor anti-parallel pseudospin configurations.
 
\begin{figure}[]
    \centering
    \includegraphics[width=\linewidth]{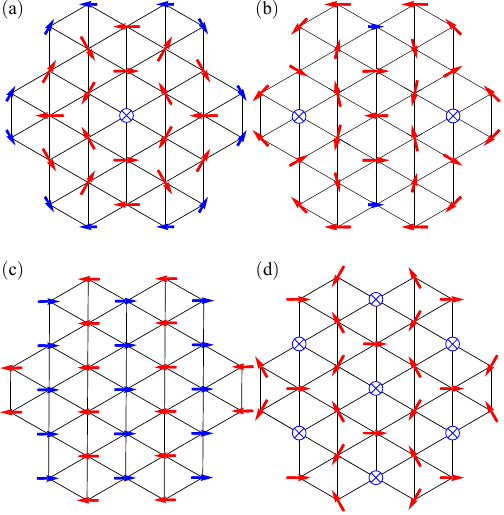}
    \caption{Multipolar ordered patterns found in the triangular lattice are shown for (a) BV, (b) 2V, (c) AFQO, and (d) SV phases.  The quadrupolar and octupolar moments are parallel and perpendicular to the triangular
 plane, respectively. Thus the length of the arrows indicate the magnitude of the quadrupole component. Red arrows represent downward-oriented octupole moment, while blue arrows represent upward-oriented octupole moments. }
    \label{fig:mesh4}
\end{figure}

\begin{figure}[!t]
    \centering
        \includegraphics[width=\linewidth]{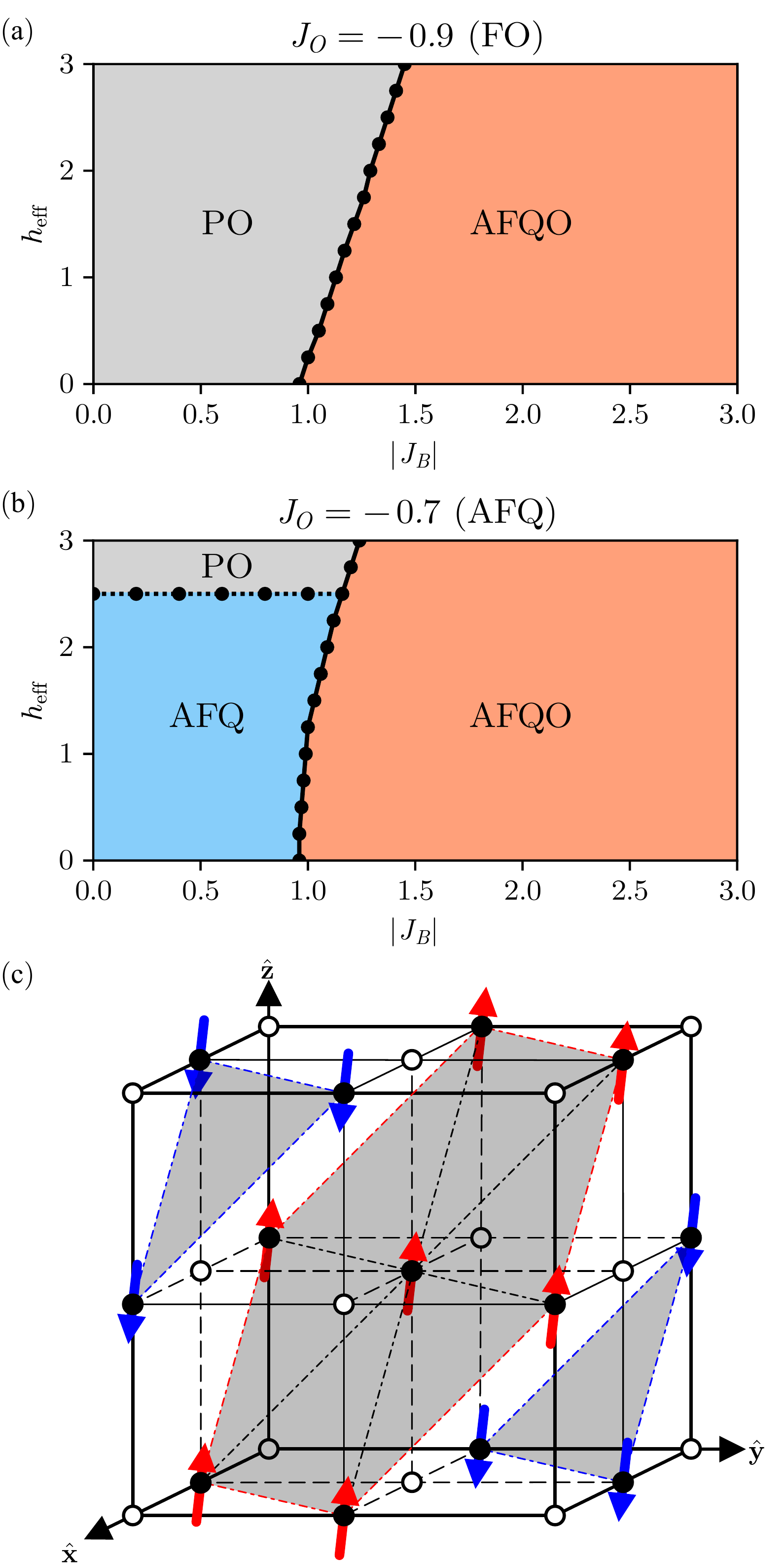}
    \caption{Classical phase diagrams computed by Monte Carlo simulated annealing on the FCC lattice when a field is applied in the $[111]$-direction and (a) $J_O = -0.9$, and (b) $J_O = -0.7$. Solid and dashed lines represent first and second order phase transitions, respectively. Note that the ferro-octupolar (FO) phase at $h_{\rm eff}=0$ in (a) becomes the polarized octupolar (PO) state at $h_{\rm eff}\neq 0$. (c) The pattern of the multipolar moments in the AFQO phase is shown: the quadrupolar and octupolar moments are parallel and perpendicular to the grey-shaded planes, respectively.}
    \label{fig:mesh5}
\end{figure}

\subsection{FCC lattice}
The classical phase diagrams obtained using Monte Carlo simulated annealing for the FCC lattice are presented in 
 Fig. 5(a) and (b) for $J_O = -0.9$ and $J_O = -0.7$, respectively. Here we utilize the same fixed value for $J_\tau$ and $J_Q$ as for the triangular lattice. 
 
 First considering $J_O = -0.9$, we observe two phases. Starting from the FO phase, increasing the magnitude of the applied field does not induces a phase transition, as expected since the effective field is parallel to the octupolar moment direction of FO. However increasing $J_B$ at any fixed $h_{\rm eff}$ induces a phase transition to the AFQO phase where the planes of aligned spins are perpendicular to $[\bar{1}\bar{1}1]$, as shown in Fig. 5(c). Analogously to the triangular lattice, in the FCC lattice, the AFQO configuration arises when large $J_B$ dominates the Hamiltonian, as the $J_B$ term promotes this specific configuration.
Similarity, taking $J_O = -0.7$ and starting in the AFQ phase, we see similar behavior to the triangular lattice for small $J_B$, where increasing $h_\text{\rm eff}$ results in octupolar component of pesduespin increasing in magnitude until a second order phase transition into the FO state is observed at $h_{\rm eff} = 2.4$. 

Due to the cancellation of the in-plane and out-of-plane $J_B$ terms with opposite signs, the intriguing vortex phases disappear in the FCC lattice, implying that the vortex phases are due to the field-induced bond-dependent $J_B$ exchange interaction. 
Lastly, we note that increasing $J_B$ incudes the AFQO phase.
This is the most interesting feature of the phase diagram of the FCC lattice as the emergence of the AFQO phase occurs via the bond-dependent $J_B$ exchange interaction enabled by the external field along the octupolar moment direction.

\section{Discussion and Summary}

 It was noted that the non-Kramer $E_g$ doublet is the lowest state of $5d^2$ ions when the Hund's coupling, SOC, and mixture of $t_{2g}$ and $e_g$ orbitals are considered, which carries quadrupolar and octupolar moments. However, the effects of the magnetic field beyond the effective Zeeman field have been less explored: the effects of the magnetic field on multipolar systems can be understood through the symmetry analysis. For example, when the field is aligned along the $[111]$-axis, which corresponds to the direction of the octupolar moment, the octupolar moment ($s^c$) couples to the cubic order of the field. This coupling takes the form $ (h_x h_y h_z) s^c/\Delta^2$ where $\Delta$ is the energy splitting between the double $E_g$ and triplet $T_{2g}$, as allowed by symmetry.  On the other hand, the field applied along $[1\bar{1}0]$-axis allows the coupling with the quadrupole moment ($s^b$), which is expressed as $ (h_x^2 - h_y^2) s^b/\Delta$. They act as an effective Zeeman field for the multipolar moments. Rayyan et al\cite{Rayyan2023PRB} have recently noted in a honeycomb lattice that the magnetic field allows the coupling between the doublet and triplet leading to another bond-dependent interaction denoted by $J_B$, which is linear in $h$. Thus, in small field regions $h < \Delta$, the $J_B$ term can dominate over the effective Zeeman field. 

We have demonstrated that the field-induced bond-dependent $J_B$ exchange interaction gives rise to rich phase diagrams in triangular and FCC lattice multipolar systems. The magnetic field along the [111]-axis turns the FO phase to an AFQO phase in the FCC lattice relevant to Os double perovskites. However, it is challenging to achieve the size of $J_B$ to be $1$ meV in Os double perovskites. Further studies are needed to quantify the strength of $J_B$ beyond the second order in strong coupling expansion. Investigating other phase space with antiferro-octupolar interaction, i.e., $J_O >0$ presents another intriguing direction: the antiferromagnetic Ising interaction on the triangular and FCC lattices is frustrated, making the effects of $J_B$ interesting subjects for further investigation.  Exploring $5d^2$ triangular lattices such as HfCl$_2$ would be another interesting direction for future studies. One might be concerned that $5d^2$ physics could be disrupted by a trigonal distortion. However, it has been shown that the $E_g$ doublet remains the lowest-energy state even in the presence of a small negative trigonal field \cite{Rayyan2023PRB2}. The vortex phases found in the triangular lattice, both with and without an external magnetic field, are particularly interesting for future research. 

In summary, the effects of the magnetic field on mulitpolar systems composed of quadrupolar and ocupolar moments in the triangular and FCC lattices are investigated. Intriguing vortex phases are found in the triangular lattice due to the field-induced bond-dependent interaction enabled by the external magnetic field. In the FCC lattice, the magnetic field along the $c$-axis parallel to $s^c$ (octupolar moment operator) leads to the antiferro-quadurole and antiferro-octupole phase. This is unexpected, since the effective Zeeman term along the $c$-axis favors the polarization of the octupolar moment, the transition to the FO occurs if the ground state is AFQ, or remains FO (or polarized octupolar state which is connected to the FO) if the ground state is FO phase to begin with. However, due to the bond-dependent interaction $J_B$ which is linear in $h$ leads to antiferromagnetically ordered quadrupolar moments in the FCC lattice. 

\section*{Acknowledgments}
This work is supported by the NSERC Discovery Grant No. 2022-04601. H.Y.K acknowledges support from the Canada Research Chairs Program and 
the Simons Emmy Noether fellowship of the Perimeter Institute,
supported by a grant from the Simons Foundation (1034867, Dittrich).
Computations were performed on the Niagara supercomputer at
the SciNet HPC Consortium. SciNet is funded by: the
Canada Foundation for Innovation under the auspices of
Compute Canada; the Government of Ontario; Ontario
Research Fund - Research Excellence; and the University
of Toronto.

\bibliographystyle{unsrt}
\bibliography{bibliography}

\end{document}